\documentclass[runningheads]{llncs}
\usepackage{graphicx}
\usepackage{setspace}
\usepackage{amsmath}
\usepackage{multirow}
\begin{document}
\title{VM-UNET-V2: Rethinking Vision Mamba UNet for Medical Image Segmentation  }
%
%
\author{Mingya Zhang\inst{1} \and
Yue Yu\inst{2} 
Limei Gu\inst{3} Tingsheng Lin\inst{3} and Xianping Tao\inst{1}}
\authorrunning{F. Author et al.}
%
\institute{State Key Laboratory for Novel Software Technology, Nanjing University
\email{dg20330034@smail.nju.edu.cn}\\
 \and
Huazhong University of Science and Technology\\
 \and Jiangsu Province Hospital of Chinese Medicine}
\maketitle              
\begin{abstract}
In the field of medical image segmentation, models based on both CNN and Transformer have been thoroughly investigated. However, CNNs have limited modeling capabilities for long-range dependencies, making it challenging to exploit the semantic information within images fully. On the other hand, the quadratic computational complexity poses a challenge for Transformers.
Recently, State Space Models (SSMs), such as Mamba, have been recognized as a promising method. They not only demonstrate superior performance in modeling long-range interactions, but also preserve a linear computational complexity. Inspired by the Mamba architecture, We proposed Vison Mamba-UNetV2, the Visual State Space (VSS) Block is introduced to capture extensive contextual information, the Semantics and Detail Infusion (SDI) is introduced to augment the infusion of low-level and high-level features.
We conduct comprehensive experiments on the ISIC17, ISIC18, CVC-300, CVC-ClinicDB, Kvasir, CVC-ColonDB and ETIS-LaribPolypDB public datasets.
The results indicate that VM-UNetV2 exhibits competitive performance in medical image segmentation tasks.
Our code is available at https://github.com/nobodyplayer1/VM-UNetV2.

\end{abstract}
\section{Introduction}
As medical imaging technology continues to advance, medical images have become a crucial tool for diagnosing diseases and planning treatments~\cite{cheng2016computer}. Among the fundamental and critical techniques in medical image analysis, medical image segmentation holds a significant place. This process involves distinguishing pixels of organs or lesions in medical images, such as CT scans~\cite{golan2016lung} and Endoscopy~\cite{tang2020development} videos. Medical image segmentation is one of the most difficult tasks in medical image analysis, with the goal of providing and extracting vital information regarding the shape and volume of these organs or tissues.
Deep learning techniques have been used recently to improve medical image segmentation. These models extract useful information from images, increase accuracy, and adapt to different datasets and tasks.

A common approach for semantic image segmentation is the use of an Encoder-Decoder network with skip connections. In this framework, the Encoder captures hierarchical and abstract features from an input image. The Decoder, on the other hand, uses the feature maps produced by the Encoder to build a pixel-wise segmentation mask or map, attributing a class label to each pixel in the input image. Numerous studies have been done to integrate global information into feature maps and enhance multi-scale features, leading to significant enhancements in segmentation performance~\cite{chen2021transunet,zhou2018unet++,liu2018path,chen2018encoder,woo2018cbam}.

U-Net~\cite{ronneberger2015u} is a pivotal architecture celebrated for its balanced Encoder-Decoder design and the incorporation of skip connections. This structure allows for the extraction of feature information at multiple levels through its various encoders and decoders. Furthermore, skip connections facilitate the effective transition of this feature information.
Numerous studies on U-Net primarily focus on the following aspects:
Encoder section - Different backbones are replaced to obtain feature maps of varying levels~\cite{woo2018cbam};
Skip connections - Various channel attention mechanisms are adopted, and different connection parts are interchanged~\cite{peng2023u};
Decoder section - Different sampling and feature fusion schemes are used~\cite{yu2015multi,zhou2020acnn}.

CNN-based models, due to their local receptive field, struggle to capture long-range information, which can lead to poor feature extraction and suboptimal segmentation results. Transformer-based models excel in global modeling but their self-attention mechanism's quadratic complexity creates high computational costs, especially in tasks like medical image segmentation that require dense predictions. These limitations necessitate a new architecture for medical image segmentation that can capture long-range information efficiently while maintaining linear computational complexity.
Recent advancements in State Space Models (SSMs), particularly Structured SSMs (S4), provide an effective solution due to their proficiency in handling long sequences. e.g., Mamba~\cite{gu2023mamba}. The Mamba model augments S4 with a selective mechanism and hardware optimization, demonstrating outstanding performance in dense data domains.
The incorporation of the Cross-Scan Module (CSM) in the Visual State Space Model (VMamba)~\cite{liu2024vmamba} further boosts Mamba's suitability for computer vision tasks. It does this by facilitating the traversal of the spatial domain and transforming non-causal visual images into ordered patch sequences.

Influenced by the success of VMamba~\cite{liu2024vmamba} in image classification task and VM-Unet~\cite{ruan2024vm} in medical image segmentation.  
Following the framework of UNetV2~\cite{peng2023u}, this paper introduces the Vision Mamba UNetV2(VM-UNetV2), we re-integrate low-level and high-level features, infusing semantic information into the low-level features, while using more detailed information to refine the high-level features.

We carry out exhaustive experiments on tasks related to gastroenterology semantic segmentation task and skin lesion segmentation to showcase the capabilities of pure SSM-based models in the field of medical image segmentation. 
In particular, we perform extensive testing on the ISIC17, ISIC18, CVC-300, CVC-ClinicDB, Kvasir, CVC-ColonDB and ETIS-LaribPolypDB public datasets and ZD-LCI-EGGIM our private dataset. The outcomes suggest that VM-UNetV2 can deliver competitive results.

The primary contributions of this study can be encapsulated in the following points: 1) We proposed VM-UnetV2, We have pioneered the exploration of better SSM-based algorithms in medical image segmentation.
2) Exhaustive experiments are performed on seven datasets, with outcomes demonstrating that VM-UNetV2 showcases significant competitiveness.
3) We are pioneering the exploration of combining SSM-based with Unet variants, driving the development of more efficient and effective SSM-based segmentation algorithms.

The rest of the paper is organized as follows: Sec.\ref{sec_methods} introduces the algorithm design consideration. Sec.\ref{sec_experiments} presents the experiments details and results. Sec.\ref{sec_conclusion} concludes the paper.

\section{Methods}\label{sec_methods}
\subsection{Preliminaries}

In contemporary SSM-based models, namely, Structured State Space Sequence Models (S4) and Mamba~\cite{gu2023mamba,liu2024vmamba,ruan2024vm}, both depend on a traditional continuous system that maps a one-dimensional input function or sequence, represented as $x\left ( t \right ) \in  R $, through intermediary implicit states $h\left ( t \right ) \in  R^{N}$ to an output $y\left ( t \right ) \in  R $. This process can be depicted as a linear Ordinary Differential Equation (ODE):

\begin{equation}
\begin{split}
         &h{}'\left ( t \right )  =Ah\left ( t \right ) +Bx\left ( t \right ) \\
         &y\left ( t \right ) =Ch\left ( t \right )
\end{split}
\end{equation}

where $\mathbf{A} \in R^{N \times N} $ represents the state matrix, while $\mathbf{B} \in R^{N \times 1} $ and $\mathbf{C} \in R^{N \times 1} $ denote the projection parameters.

S4 and Mamba discretize this continuous system to adapt it better for deep learning contexts. Specifically, they incorporate a timescale parameter $\boldsymbol{\Delta}$ and convert $\mathbf{A}$ and $\mathbf{B}$ into discrete parameters  $\mathbf{\overline{A}}$ and $\mathbf{\overline{B}}$ using a consistent discretization rule. The zero-order hold (ZOH) is typically utilized as the discretization rule and can be outlined as follows:

\begin{equation}
\begin{split}
        &\overline
{\mathbf{A}}=\exp (\boldsymbol{\Delta} \mathbf{A}) \\
        &\overline{\mathbf{B}}=(\boldsymbol{\Delta} \mathbf{A})^{-1}(\exp (\boldsymbol{\Delta} \mathbf{A})-\mathbf{I}) \cdot \boldsymbol{\Delta} \mathbf{B}
\end{split}
\end{equation}

Following discretization, SSM-based models can be calculated in two distinct methods: linear recurrence or global convolution, which are denoted as equations \eqref{eq:1} and \eqref{eq:2}, respectively.

\begin{equation}
    \begin{split}
        &h{}' \left ( t \right ) = \overline{\mathbf{A}}  h\left ( t \right ) + \overline{\mathbf{B}}x\left ( t \right ) \\
        &y\left ( t \right ) =\mathbf{C}h\left ( t \right )  \label{eq:1}
    \end{split}
\end{equation}

\begin{equation}
    \begin{split}
        & \overline{K} = \left ( \mathbf{C}\overline{\mathbf{B}} , \mathbf{C}\overline{\mathbf{A}\mathbf{B}} ,...,\mathbf{C}\overline{\mathbf{A} } ^{L-1}\overline{\mathbf{B}}    \right )  \\
        & y=x*\overline{\mathbf{K} } \label{eq:2}
    \end{split}
\end{equation}

where $\overline{\mathbf{K} } \in R^{L} $ represents a structured convolutional kernel, and $L$ denotes the length of the input sequence $x$.

\subsection{VM-UNetV2 Architecture}

\begin{figure}
    \centering
    \includegraphics[width=\textwidth]{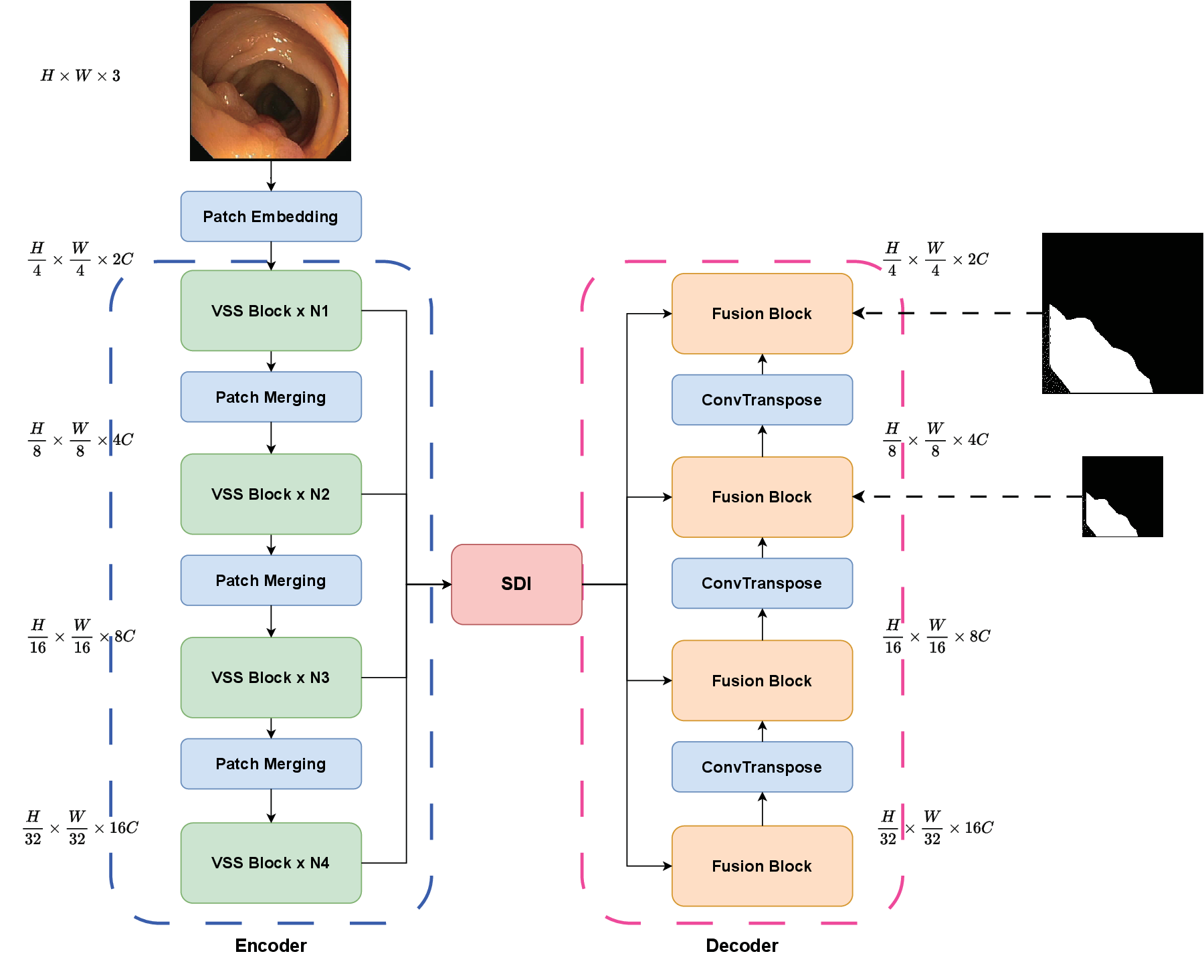}
    \caption{The overall architecture of Vision Mamba UNetV2 model, which consists of an Encoder module, SDI(semantics and detail infusion)~\cite{peng2023u} module, and an Decoder module}
    \label{fig:structure}
\end{figure}

The comprehensive structure of Vision Mamba UNetV2 is depicted in Fig~\ref{fig:structure}. It consists of three primary modules: the Encoder, the SDI (Semantic and Detail Infusion)~\cite{peng2023u} module, and the Decoder. Given an input image I, where $I\in{R^{H \times W \times 3}}$,
the encoder generates features at M levels. We represent the features at the $i_{th}$ level as $f_{i}^{o}$, where $1\le i \le M$. These accumulated features,$\left \{ f_{1}^{o}, f_{2}^{o},...,f_{M}^{o} \right \} $ are subsequently forwarded to the SDI module for further enhancement.

As described the Figure, the Encoder output channel of $f_{i}$ is $2^
{i}\times C $, $\left \{ f_{1}^{o}, f_{2}^{o},...,f_{M}^{o} \right \} $ collectively enter the SDI module for feature fusion, and $f_{i}$ corresponds to $f{_{i}}' $ as the output of the $i_{th}$ stage. The feature of $f{_{i}}' $ is $\frac{H}{2^{i+1}} \times\frac{W}{2^{i+1}}\times 2^{i}C  $. In our model, we use deep supervision to calculate the loss of $f{_{i}}'$ and $f{_{i-1}}'$ features.\par
In this paper, we employ $\left [ N_{1}, N_{2}, N_{3}, N_{4} \right ] $ VSS blocks across Encoder four stages, with the channel counts for each stage being $\left [ C, 2C, 4C, 8C \right ] $.
From our observations in VMamba~\cite{liu2024vmamba}, the different values of $N_{3}$ and $c$ are important factors that distinguish between Tiny, Small, and Base framework specs. Following the specifications of VMamba, we let C take the value of 96, $N_{1}$ and $N_{2}$ each take the value of 2, and $N_{3}$ take values from the set [2,9,27]. This represents our intention to use the Tiny and Small models of VMamba as the backbone for our ablation experiments.

\subsection{VSS And SDI Block}

\begin{figure}
    \centering
    \includegraphics[width=\textwidth]{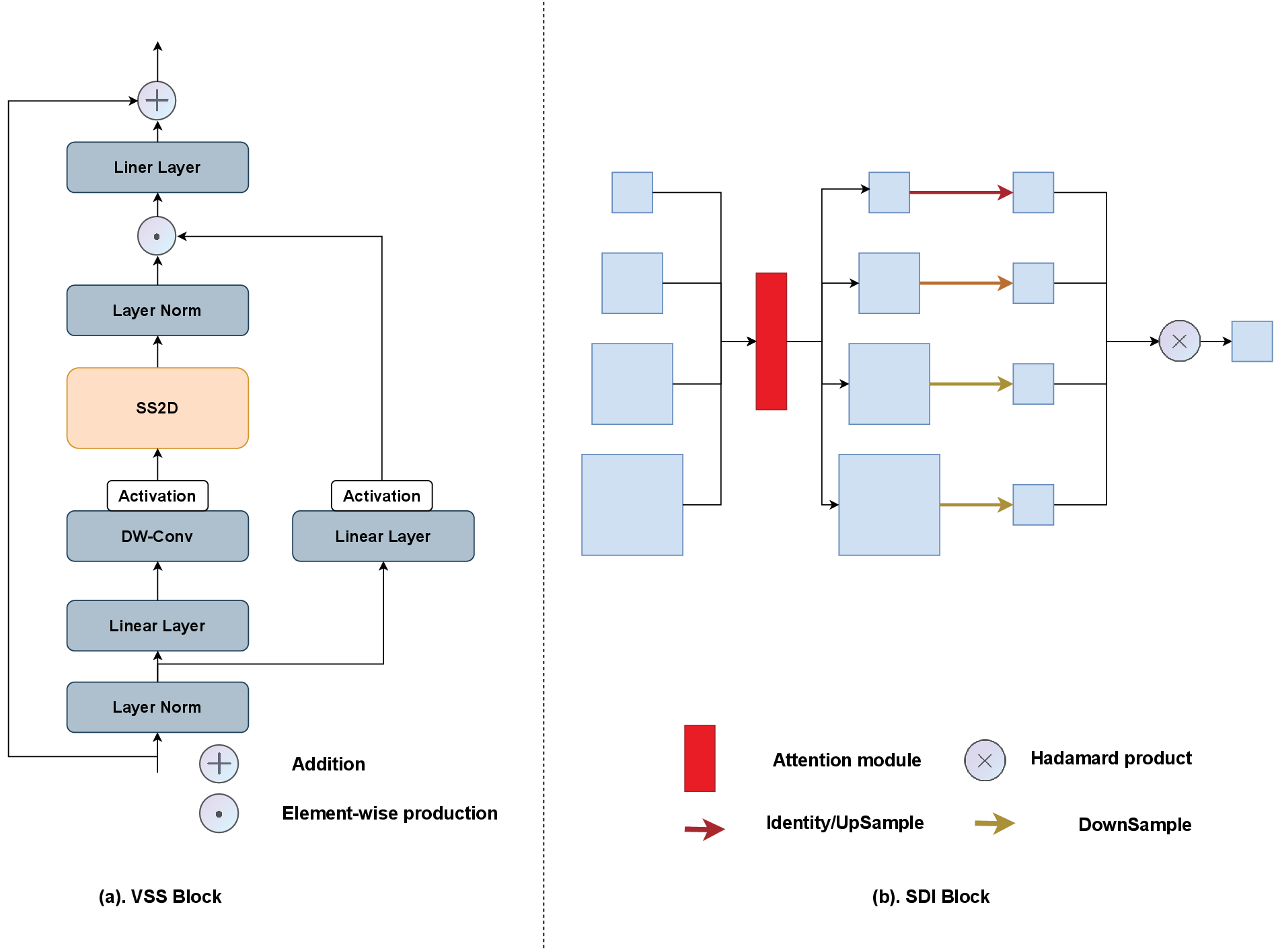}
    \caption{a. VSS Block as the backbone of VMUNetV2, and the SS2D is the core of VSS block b.( Semantics and Detail Infusion)SDI module consists of backbone's output features, an Attention module, output features of SDI are with the different size the same as backbone's output features. }
    \label{fig:vss_sdi}
\end{figure}

The VSS block derived from VMamba, As the backbone part of the VM-UNetV2 Encoder, The structure of VSS block is illustrated in Figure~\ref{fig:vss_sdi}$\left( a\right)$.
The input is first processed through an initial linear embedding layer, after which it divides into two separate information streams. One stream is directed through a $3\times3$ depth-wise convolution~\cite{howard2017mobilenets} layer and subsequently a Silu activation~\cite{shazeer2020glu} function before it enters the main 2D-Selective-Scan(SS2D) module. The SS2D output then proceeds through a layer normalization layer and is combined with the output from the other information stream, which has also been processed through a Siluactivation. This merged output constitutes the final result of the VSS block.

The SDI module~\cite{peng2023u}, As described in Figure~\ref{fig:vss_sdi} (b). With the hierarchical feature maps  $f_{i}^{0} = \frac{H}{2^{i+1}} \times\frac{W}{2^{i+1}}\times 2^{i}C, 1\le i\le 4 $ generated by the encoder, where $i$ represents the $i_{th}$ level.

Different Attention mechanisms can be used in the SDI module to calculate the attention scores for both space and channel. Following what is mentioned in UNetV2~\cite{peng2023u}, we use CBAM~\cite{woo2018cbam} to implement spatial and temporal attention. The calculation formula is as follows, and $\phi _{i}^{att}$ represents the $i_{th}$ attention calculation:
\begin{equation}
    f_{i}^{1}=\phi _{i}^{att}\left ( f_{i}^{0} \right )
\end{equation}
Then we use $1\times 1$ convolution to align the channel of $f_{i}^{1}$ to $c$ the resulted feature map is denoted as $f_{i}^{2} \in R^{H_{i}\times W_{i} \times c}$.\par
In SDI decoder $i_{th}$ stage, $f_{i}^{2}$ denotes target reference.Then we adjust the sizes of the feature maps at every $j_{th}$ level to match the size of $f_{i}^{2}$, as formulated below:

\begin{equation}
    f_{i j}^{3}= \left\{\begin{array}{ll}
\mathrm{G_{d}}\left(f_{j}^{2},\left(H_{i}, W_{i}\right)\right) & \text { if } j<i, \\
\mathrm{G_{I}}\left(f_{j}^{2}\right) & \text { if } j=i, \\
\mathrm{G_{U}}\left(f_{j}^{2},\left(H_{i}, W_{i}\right)\right) & \text { if } j>i,
\end{array}\right. \label{GG}
\end{equation}
In the formula~\ref{GG}, $\mathrm{G_{d}}$, $\mathrm{G_{i}}$ and $\mathrm{G_{u}}$ represent adaptive average pooling, identity mapping, and bilinearly interpolating. In the formula~\ref{GG2},
$\theta_{ij}$ represents the parameters of the smooth convolution, and
$f_{ij}^{4}$is the $j_{th}$ smoothed feature map at the $i_{th}$ level.
Here, $H\left (   \right ) $ represents the Hadamard product. Subsequently, $f_{i}^{5}$ is forwarded to the decoder at the $i_{th}$ level for further reconstruction of resolution and segmentation.
\begin{equation}
    \begin{split}
        &f_{ij}^{4}=\theta _{ij}\left ( f_{ij}^{3} \right )  \\
        &f_{i}^{5}=H\left ( \left [f_{i1}^{4}, f_{i2}^{4},f_{i3}^{4}, f_{i4}^{4} \right ]  \right ) \label{GG2}
    \end{split}
\end{equation}

\subsection{Loss function}

For our medical image segmentation tasks, we primarily employ basic Cross-Entropy and Dice loss as the loss function cause all of our dataset masks comprise two classes, which are a singular target and the background.

\begin{equation}
    \begin{split}
        &L_{\text {BceDice }}=\lambda_{1} L_{\mathrm{Bce}}+\lambda_{2} L_{\text {Dice }} \\
        &L_{Bce} = -\frac{1}{N}\sum_{1}^{N}\left [ y_{i} log\left (  \hat{y}_{i} \right ) +\left (  1-y_{i}\right )log\left ( 1-\hat{y}_{i}  \right )   \right ] \\
        &L_{\text {Dice }}=1-\frac{2|X \cap Y|}{|X|+|Y|}
    \end{split}
\end{equation}

$\left ( \lambda_{1}, \lambda_{2} \right ) $ are constants, with $\left ( 1, 1 \right ) $ often selected as the default parameters.

\section{Experiments and results}\label{sec_experiments}

\subsection{Datasets}




We use three types of datasets to verify the effectiveness of our framework. The first type is the open-source skin diseases dataset, including ISIC 2017 and ISIC 2018, we split the skin datasets in a 7: 3 ratio for use as training and testing sets.
The second is the open-source gastrointestinal polyp dataset, which includes Kvasir-SEG, ClinicDB, ColonDB, Endoscene, and ETIS, in this type datasets we follow the experimental setups in PraNet.
For these datasets, we provide detailed evaluations on several metrics, including Mean Intersection over Union(mIoU), Dice Similarity Coefficient(DSC), Accuracy(Acc), Sensitivity(Sen), and Specificity(Spe).

\subsection{Experimental setup}

Following the VMamba work, we adjust the image dimensions in all datasets to $256 \times 256$ pixels. With the aim of curbing overfitting, we also bring in data augmentation methods, such as random flipping and random rotation. 
In terms of operational parameters, we have the batch size set at 80, with the AdamW optimizer engaged starting with a learning rate of 1e-3. We make use of CosineAnnealingLR as the scheduler, with its operation spanning a maximum of 50 iterations and the learning rate going as low as 1e-5. We conduct our training over the course of 300 epochs.
For the VM-UNetV2, the encoder units' weights are initially set to align with those of VMamba-S.
The implementation was carried out on an Ubuntu 20.04 system, using Python3.9.12, PyTorch2.0.1, and CUDA11.7, All experiments are conducted on a single NVIDIA RTX V100 GPU.

\subsection{Results}

We compare VM-UNetV2 with some state-of-the-art models, presenting the experimental results in Table~\ref{tab:isic_all} and Table~\ref{tab:polyp_all}.
For the ISIC datasets, our VM-UNetV2 outperforms other models in terms of the mIoU, DSC and Acc metrics. 
In the polyp-related datasets, our model also surpasses the state-of-the-art model UNetV2 in all metrics, with an increase of up to $7\%$ in the mIoU parameter.

\begin{table}[]
\caption{Comparative experimental results on the ISIC17 and ISIC18 datasets(Bold indicates the best)}
\label{tab:isic_all}
\begin{tabular}{ccccccc}
\hline
\textbf{Dataset}         & \textbf{Model}     & \textbf{mloU(\%)↑} & \textbf{DSC(\%)↑} & \textbf{Acc(\%)↑} & \textbf{Spe(\%)↑} & \textbf{Sen(\%)↑} \\ \hline
\multirow{7}{*}{ISIC17}  & UNet~\cite{ronneberger2015u}               & 76.98              & 86.99             & 95.65             & 97.43             & 86.82             \\
                         & UTNetV2~\cite{gao2022multi}            & 77.35              & 87.23             & 95.84             & 98.05             & 84.85             \\
                         & TransFuse~\cite{zhang2021transfuse}          & 79.21              & 88.40             & 96.17             & 97.98             & 87.14             \\
                         & MALUNet~\cite{ruan2022malunet}            & 78.78              & 88.13             & 96.18             & 98.47             & 84.78             \\
                         & UNetV2~\cite{peng2023u}             & 82.18              & 90.22             & 96.78             & 98.40             & 88.71             \\
                         & VM-UNet~\cite{ruan2024vm}            & 80.23              & 89.03             & 96.29             & 97.58             & 89.90             \\
                         & \textbf{VM-UNetV2} & \textbf{82.34}     & \textbf{90.31}    & \textbf{96.70}    & 97.67             & \textbf{91.89}    \\ \hline
\multirow{10}{*}{ISIC18} & UNet~\cite{ronneberger2015u}               & 77.86              & 87.55             & 94.05             & 96.69             & 85.86             \\
                         & UNet++~\cite{zhou2018unet++}             & 78.31              & 87.83             & 94.02             & 95.75             & 88.65             \\
                         & Att-Unet~\cite{oktay2018attention}           & 78.43              & 87.91             & 94.13             & 96.23             & 87.60             \\
                         & UTNetV2~\cite{gao2022multi}            & 78.97              & 88.25             & 94.32             & 96.48             & 87.60             \\
                         & SANet~\cite{wei2021shallow}              & 79.52              & 88.59             & 94.39             & 95.97             & 89.46             \\
                         & TransFuse~\cite{zhang2021transfuse}          & 80.63              & 89.27             & 94.66             & 95.74             & 91.28             \\
                         & MALUNet~\cite{ruan2022malunet}            & 80.25              & 89.04             & 94.62             & 96.19             & 89.74             \\
                         & UNetV2~\cite{peng2023u}             & 80.71              & 89.32             & 94.86             & 96.94             & 88.34             \\
                         & VM-UNet~\cite{ruan2024vm}            & 81.35              & 89.71             & 94.91             & 96.13             & 91.12             \\
                         & \textbf{VM-UNetV2} & \textbf{81.37}     & \textbf{89.73}    & \textbf{95.06}    & \textbf{97.13}    & 88.64             \\ \hline
\end{tabular}
\end{table}

\begin{table}[]
\caption{Comparative experimental results on the Kvasir-SEG, ClinicDB, ColonDB, ETIS and CVC-300 datasets(Bold indicates the best)}
\label{tab:polyp_all}
\begin{tabular}{lllllll}
\hline
\textbf{Dataset}            & \textbf{Model}    & \textbf{mloU(\%)↑} & \textbf{DSC(\%)↑} & \textbf{Acc(\%)↑} & \textbf{Spe(\%)↑} & \textbf{Sen(\%)↑} \\ \hline
\multirow{3}{*}{Kvasir-SEG} & UNetV2            & 84.00              & 91.30             & 97.47             & 99.08             & \textbf{88.39}    \\
                            & VMUnet            & 80.32              & 89.09             & 96.80             & 98.49             & 87.21             \\
                            & \textbf{VMUnetV2} & \textbf{84.15}     & \textbf{91.34}    & \textbf{97.52}    & \textbf{99.25}    & 87.71             \\ \hline
\multirow{3}{*}{ClinicDB}   & UNetV2            & 83.85              & 91.21             & 98.59             & 99.16             & 91.99             \\
                            & VMUnet            & 81.95              & 90.08             & 98.42             & 99.18             & 89.73             \\
                            & \textbf{VMUnetV2} & \textbf{89.31}     & \textbf{94.35}    & \textbf{99.09}    & \textbf{99.38}    & \textbf{95.64}    \\ \hline
\multirow{3}{*}{ColonDB}    & UNetV2            & 57.29              & 72.85             & 96.19             & 98.43             & 68.46             \\
                            & VMUnet            & 55.28              & 71.20             & 96.02             & 98.45             & 65.89             \\
                            & \textbf{VMUnetV2} & \textbf{60.98}     & \textbf{75.76}    & \textbf{96.54}    & \textbf{98.46}    & \textbf{72.68}    \\ \hline
\multirow{3}{*}{ETIS}       & UNetV2            & 71.90              & 83.65             & 98.35             & 98.61             & \textbf{92.96}    \\
                            & VMUnet            & 66.41              & 79.81             & 98.26             & 99.33             & 75.79             \\
                            & \textbf{VMUnetV2} & \textbf{72.29}     & \textbf{83.92}    & \textbf{98.47}    & \textbf{98.96}    & 88.07             \\ \hline
\multirow{3}{*}{CVC-300}    & UNetV2            & 82.86              & 90.63             & 99.34             & 99.54             & 93.82             \\
                            & VMUnet            & 79.55              & 88.61             & 99.20             & 99.44             & 92.46             \\
                            & \textbf{VMUnetV2} & \textbf{89.31}     & \textbf{94.35}    & 99.08             & 99.38             & \textbf{95.64}    \\ \hline
\end{tabular}
\end{table}

Beyond evaluating the accuracy of the models, we also assess the computational complexity of different models, leveraging the advantage of VMamba's linear complexity. We use metrics such as model inference speed FPS (frames per second), model parameter count (Params), and the number of floating point operations (FLOPs) for this evaluation, as depicted in Table~\ref{tab:metirc_model}.
$(3, 256, 256)$ represents the size of the input image. All the tests are conducted on an NVIDIA V100 GPU. the FLOPs and FPS of VM-UNetV2 are superior to those of others.

\begin{table}[]
\centering
\caption{Comparison of computational complexity, GPU memory usage, and inference time, using an NVIDIA V100 GPU(Bold indicates the best).}
\label{tab:metirc_model}
\begin{tabular}{ccccc}
\hline
\textbf{Model} & \textbf{Input size} & \textbf{Params(M)↓} & \textbf{FLOPs(G)↓} & \textbf{FPS↑}  \\ \hline
UNetV2         & (3, 256, 256)       & 25.15               & 5.40               & \textbf{32.74} \\
VM-Unet        & (3, 256, 256)       & 34.62               & 7.56               & 20.612         \\
VM-UnetV2      & (3, 256, 256)       & \textbf{17.91}      & \textbf{4.40}      & 32.58          \\ \hline
\end{tabular}
\end{table}

\subsection{Ablation studies}
In this section, we conduct ablation experiments on the initialization of VM-UNetV2 Encoder and the Deep Supervision operation of Decoder using the polyp datasets.
As stated in the VMamba~\cite{liu2024vmamba} paper, the depth of the Encoder and the number of channels in the feature map determine the scale of VMamba. In this paper, the proposed VM-UNetV2 only uses the pre-trained weights of VMamba on ImageNet-1k for the Encoder part. Therefore, when conducting the model scale ablation experiment in this study, we only vary the depth of the Encoder, as shown in the Table~\ref{tab:ablation}. For the output features, we employ a Deep Supervision mechanism, using a fusion of two layers of output features, which are then compared with the real labels for loss computation.

As shown in Table~\ref{tab:ablation} and Table~\ref{tab:ablation2}, when the depth of the Encoder is set to $\left [ 2,2,9,2 \right ] $, the segmentation evaluation metrics are relatively better. Therefore, when using VM-UNetV2, there is no need to choose a particularly large depth. In most cases where the Deep Supervision mechanism is used, the segmentation evaluation metrics are relatively better, but it is not a decisive factor. For different datasets, ablation experiments need to be conducted separately to determine whether to adopt the Deep Supervision mechanism.

\begin{table}[]
\caption{Ablation studies on Encoder Depth and Deep Supervision of VM-UNet-V2 (part 1)}
\label{tab:ablation}
\begin{tabular}{cccccc}
\hline
\multirow{2}{*}{Encoder Depth}  & \multirow{2}{*}{Deep Supervision} & Kvasir-SEG &          & \multicolumn{2}{c}{ClinicDB} \\ \cline{3-6} 
                                &                                   & mloU(\%)↑  & DSC(\%)↑ & mloU(\%)↑     & DSC(\%)↑     \\ \hline
\multirow{2}{*}{{[}2,2,2,2{]}}  & TRUE                              & 82.93      & 90.67    & 83.08         & 90.76        \\
                                & FALSE                             & 82.59      & 90.47    & 85.40         & 92.13        \\ \hline
\multirow{2}{*}{{[}2,2,9,2{]}}  & TRUE                              & 85.23      & 92.03    & 89.02         & 94.19           \\
                                & FALSE                             & 84.15      & 91.39    & 89.31         & 94.35        \\ \hline
\multirow{2}{*}{{[}2,2,27,2{]}} & TRUE                              & 82.90      & 90.65    & 88.45         & 93.87        \\
                                & FALSE                             & 79.57      & 88.63    & 80.58         & 89.24        \\ \hline
\end{tabular}
\end{table}

\begin{table}[]
\centering
\caption{Ablation studies on Encoder Depth and Deep Supervision of VM-UNet-V2 (part 2)}
\label{tab:ablation2}
\begin{tabular}{cccccc}
\hline
\multicolumn{2}{c}{ColonDB} & \multicolumn{2}{c}{ETIS} & \multicolumn{2}{c}{CVC-300} \\ \hline
mloU(\%)↑     & DSC(\%)↑    & mloU(\%)↑   & DSC(\%)↑   & mloU(\%)↑     & DSC(\%)↑    \\ \hline
57.68         & 73.16       & 66.22       & 79.68      & 76.50         & 86.68       \\
54.44         & 70.5        & 68.05       & 80.98      & 77.66         & 87.42       \\ \hline
62.94         & 77.26       & 67.94       & 80.91      & 80.43         & 89.15          \\
60.98         & 75.76       & 72.29       & 83.92      & 79.23         & 88.41       \\ \hline
64.06         & 78.09       & 70.72       & 82.85      & 80.32         & 89.09       \\
57.03         & 72.63       & 63.55       & 77.72      & 77.66         & 87.42       \\ \hline
\end{tabular}
\end{table}

\section{Conclusions}\label{sec_conclusion}

In this paper, we propose a SSM-based UNet type medical image segmentation model, VM-UNetV2, which fully utilizes the capabilities of SSM-based models. We use VSS blocks and SDI to process the Encoder and Skip connection, respectively. The pre-trained weights of VMamba are used to initialize the Encoder part of VM-UNetV2, and a Deep Supervision mechanism is employed to supervise multiple output features. Our model has been extensively tested on skin disease and polyp datasets. The results demonstrate that our model is highly competitive in segmentation tasks. Complexity analysis suggested that VM-UNetV2 is also efficient in FLOPs,Params and FPS.




%
%
%
\bibliographystyle{splncs04}
\bibliography{mybibliography}

\end{document}